1. **Title page:**

**Predictive numerical simulations of double branch stent-graft deployment in an aortic arch aneurysm.**

**L. Derycke[1,3], D. Perrin[1], F. Cochennec[3], J.-N. Albertini[2], S. Avril[1].**

1. Mines Saint-Etienne, Univ Lyon, Univ Jean Monnet, INSERM, U 1059 Sainbiose, Centre CIS, 158, cours Fauriel, F - 42023 Saint-Etienne France
2. Service de Chirurgie vasculaire, Centre Hospitalier Régional Universitaire de Saint-Etienne, avenue Albert Raimond, 42270 Saint-Priez-en-Jarez, France.
3. Assistance publique des Hôpitaux de Paris, hôpital Mondor, service de chirurgie vasculaire, 51, avenue du Maréchal-de-Lattre-de-Tassigny, 94010 Créteil, France.

**Abbreviated title for the running head**:
Simulations of stent-graft deployment in aortic arch

**Name and contact for correspondence**
DERYCKE Lucie
62 boulevard Gambetta, 94130 Nogent-sur-Marne, FRANCE
+33683629898
deryckelucie@gmail.com

## 2. Abstract and key terms:

Total endovascular repair of the aortic arch represents a promising option for patients ineligible to open surgery. Custom-made design of stent-grafts (SG), such as the Terumo Aortic® RelayBranch device (DB), requires complex preoperative measures. Accurate SG deployment is required to avoid intraoperative or postoperative complications, which is extremely challenging in the aortic arch. In that context, our aim is to develop a computational tool able to predict SG deployment in such highly complex situations. A patient-specific case is performed with complete deployment of the DB and its bridging stents in an aneurysmal aortic arch. Deviations of our simulation predictions from actual stent positions are estimated based on post-operative scan and a sensitivity analysis is performed to assess the effects of material parameters. Results show a very good agreement between simulations and post-operative scan, with especially a torsion effect, which is successfully reproduced by our simulation. Relative diameter, transverse and longitudinal deviations are of 3.2±4.0%, 2.6±2.9mm and 5.2±3.5mm respectively. Our numerical simulations show their ability to successfully predict the DB deployment in complex anatomy. The results emphasize the potential of computational simulations to assist practitioners in planning and performing complex and secure interventions.

Abbreviations:
BS: Bridging stents
CT: Computed tomography
DB: Terumo Aortic® (formerly Bolton Medical®) RelayBranch device
FEA: Finite-element analysis
SG: Stent-graft
TEVAR: Thoracic Endovascular Aneurysm Repair
VMTK: Vascular Modeling Toolkit

### 3. Introduction

Endovascular procedures, using a fabric-covered and self-expanding stent-graft (SG), are the major alternative to open surgery to prevent aneurysm rupture. In most European and US medical centers specialized in aortic aneurysm and dissection repair, endovascular procedures are now considered as the first line treatment of thoracic and abdominal aortic aneurysms. The reported experience shows significant reduction of early morbidity and mortality compared to open surgical repair [1,24,25,30,31]

The supra aortic vessels (innominate, common carotid and subclavian arteries) and the aortic arch are subject to many variations in their anatomy and their curvature, making the aortic arch a very complex part of the thoracic aorta [22,29]. Moreover, the high hemodynamic forces and the mechanical constraints present in the aortic arch limit the reliability of the endovascular repair and challenge the device durability [5,16,23,35,43]. In consequence, despite the need for cardiopulmonary bypass, hypothermic circulatory arrest and its high mortality/morbidity rate, open surgery remains the gold standard for treating aortic arch pathologies such as aneurysms or dissections [27,42]. Therefore, it can be acknowledged the aortic arch remains one of the last challenges for endovascular surgery in aortic aneurysm and dissection repair. However, due to the invasiveness of open surgery, a significant fraction of high-risk patients is ineligible to open surgery.

In the past few years, new fenestrated or branched endovascular aortic grafts have been proposed for aortic arch endovascular repair, such as: the Terumo Aortic® (formely Bolton Medical®) RelayBranch device (DB) and the Branch Arch Cook® device. These devices are custom-made and have a complex design with internal tunnels and a fenestration-landing-zone to comply with any patient specific anatomy. Then, bridging stents are inserted into the internal tunnels from the supra-aortic trunk vessels to fully separate the aneurysm from the normal blood flow. Accurate preoperative sizing based on computed tomography (CT)-scan is required to generate a custom-made design.

A very thorough additional examination of this CT-scan by the physician is mandatory to predict accurate SG deployment [26]. Nowadays, preoperative planning requires strong experience in imaging and 3D reconstruction using a dedicated workstation. In case of arterial tortuosity and angulation, as in the aortic arch, the deployment prediction is challenging and measures may often be inaccurate. Moreover, due to the length of the delivery system, the ability to precisely control device deployment is limited.

In this context, finite-element analysis (FEA) could be useful to predict the deployment of the DB device in such complex anatomies in order to assist practitioners in the clinical decision-making process and during the aortic repair. Research on this topic started less than a decade ago [9,12,21]. SG mechanical properties were characterized using bench tests and used to achieve FEA of SG deployment in virtual aortas and iliac arteries. These models highlighted the importance of boundary conditions, contacts and material properties. Moreover, they established the basic assumptions about pre-stressed conditions, arterial wall properties and interactions with surrounding tissues. Recently, more complex numerical simulations of bifurcated and fenestrated SG deployment in patient-specific models have shown the validity of the different methodologies [10,11,33,34]. Auricchio *et al* were the first to report FEA results about the deployment of a tubular SG in the thoracic aorta [3,38].

Our objective is to model computationally the deployment of the complex double-branched SG in the aortic arch. In the following, we address patient-specific numerical simulations of endovascular aneurysm repair in the aortic arch using the DB device. After a thorough description on how we extended the methodology elaborated by Perrin et al [32,33] to reach our objective, we show successful comparison between our simulations and post-operative CT scan  and we report results of a sensitivity study, focusing on friction coefficients and material properties used in the FEA.

## 4. Materials and methods

Our methodology is an extension of the methodology elaborated by our group [32,33]. We developed a specific methodology for SG deployment in patient-specific abdominal aortic aneurysms. The SG is virtually crimped and then inserted in a tubular shell modeling an idealized aortic wall. Then, a transition step, named *morphing*, is achieved: the tubular shell is computationally deformed into the pre-operative geometry of the aorta, while the SG is maintained inside the deforming tubular shell with activated contact elements. Due to these constraints, the SG undergoes deformation to fit into the patient-specific aortic geometry. Finally, material properties are assigned to the aortic wall and a static mechanical equilibrium is solved between the aorta and the device. So, the model simulates neither SG insertion in a delivery sheath nor navigation through iliac arteries and the aorta. In the current study, major extensions had to be elaborated to model the DB device and to address the aortic arch specificities. We introduce these extensions hereafter.

1. Aortic arch aneurysm modeling and *morphing*

A 74-year-old male patient, treated by DB for aortic arch aneurysm, was chosen in our clinical database after informed consent and approval of the Institutional Review Board of the Saint-Etienne University Hospital. Pre- and post-operative CT scans were available, along with complete plan of the DB device. The pre-operative and the post-operative CT scans had the following parameters: slice thickness = *2mm* and *0.7mm*, pixel size = *0.5mm x 0.5mm* and *0.8mm x 0.8mm*, respectively. The pre-operative CT scan showed a *58mm* dilatation of the aortic arch zone 0 according to the Ishimaru classification [20] .

Modeling:

The aortic lumen, from the aortic valve to the descending thoracic aorta, and the supra-aortic trunk vessels, was reconstructed using the Vascular Modeling Toolkit (VMTK) library.  As calcifications did not appear in the CT scan of the clinical case considered in our study and intra-luminal thrombus was very moderate, these components were not represented in the model. The luminal surface was meshed with triangular shell elements (SFM3D3 in Abaqus® and 1.5 mm mean edge length), resulting in a total of 32125 nodes and 64039 elements. A constant wall thickness of 1.5 mm for the aorta and 1 mm for the supra-aortic trunk was assigned.

*Morphing*:

The *morphing* algorithm [32,33] was fully adapted to the aortic arch anatomy. This required accounting for the high degree of curvature and incorporation of the supra-aortic trunk vessels.

After generating the centerlines, splines were defined to describe luminal contours. Each spline had 10 control points for each cross section.

The *morphing* algorithm was applied to the aortic arch to deform the preoperative mesh using centerlines and splines as driving key-points. Each node of the aortic surface was moved to create a tubular shell with a constant diameter. Tubular shells for each supra-aortic vessel, orthogonal to the aortic centerline, were also defined. The forward and the inverse deformation between the patient-specific geometry to the geometry made of tubular shells were computed. A detailed description of the tubular virtual shell generation was previously described [32].

### 2. Stent-grafts modeling

The Double Branch Relay® from Bolton Medical® (DB) is a custom-made device. The main body has a complex design with a large single window harboring two internal tunnels for secondary connection of supra-aortic extensions to the innominate and the left carotid common arteries. Moreover, four kinds of stent rings are sewn to the graft: standard Z stent, half stent, crown stent and flattened stent.

Geometries of the graft and stent rings were obtained from the manufacturer. The four kinds of stent rings were modelled and meshed with beam elements in Matlab® (B31 in Abaqus® and 0.3 mm mean length). A linear elastic material behavior was used, reproducing the Nitinol behavior in its austenitic phase. Nitinol remained in its austenitic phase during SG deployment as the full crimping in its delivery sheath was not considered, as previously validated on different types of SGs [12,34]. An accurate geometry of the graft was created using FreeCAD® and meshed with linear 4-node elements in Abaqus® (S4R and 0.5 mean edge length). The polyester fabric was modelled as an orthotropic elastic material [12]. The internal branches were tied to the main graft. The bridging stents (BSs) were modelled following the same process as the main body of the SG (see Fig. 3.(a)).

### 3. Pre-stress of stent-graft wires

All simulations were performed using the Abaqus/Explicit v6.14 finite element solver (Dassault Systèmes, Paris, France). During the manufacturing process, stent rings are oversized with respect to the graft diameter. Accordingly, a first FEA was performed to crimp the oversized stent rings until they are in contact with the graft. A tie constraint was assigned between the oversized stent rings and the graft at the end of this simulation step, which was achieved for the three components of the DB device (the SG and the two BSs) (see Fig. 3.(a)).

After this first step, we performed a second FEA to radially compress the BSs in two virtual cylinders by gradually reducing their diameters and applying a contact constraint and to simulate their placement in the internal tunnels (see Fig. 3.(b)).

Then, the third step was an FEA to radially compress the SG with its two BSs in a cylindrical tube adjusted to the diameter of the aorta (see Fig. 3.(c)).

A fourth step was an FEA to bring the distal end of BSs to the innominate and the left carotid arteries, which were guided in two virtual cylinders, and to bring the proximal part of BSs into contact within the tunnels. An original algorithm was developed to create and to compute the displacement of the two cylinders meshed with shell elements. A friction coefficient of 0.4 was set for the contact between the different components during this step [44] (see Fig 3.(d)). Finally, in a fifth step, the virtual cylinders were removed and the main body of the SG and its BSs were released into the tubular shape of the aorta to allow their deployment (see Fig. 3.(e)).

4. <u>Simulation methodology of stent-grafts deployment</u>

The operative report did not provide precisely the location of proximal landing zone and global rotation. Therefore, the position of the landing zone for the simulations was derived directly from the post-operative CT scan.

After simulating the deployment of the device at the correct landing zone inside the tubular shape shown in Fig. 3.(e), computed deformations were applied to each node of the tubular shell to transform it back into the pre-operative geometry (see Fig. 3.(f)). General contact was activated, including the internal surface of the tubular shell, the stent rings and the grafts. This sixth FEA imposed the SGs to fit inside a shell modeling rigidly the patient-specific geometry of the aorta. A rough friction coefficient was used to keep the proximal parts of the BSs in the internal branches. As the aortic virtual shell only had a geometrical function, there was no mechanical behavior assigned to the aortic wall at this step. Thus, the final geometry of the aorta at the end of the morphing stage was stress free. Contact between SG and the aortic wall were modeled using the default Abaqus® contact algorithm with a friction coefficient set to 0.4. Only the 2cm at the proximal end on the main SG were assigned a stronger friction coefficient to ensure no sliding at the landing zone.

Finally, in the final simulation step, the shell elements modeling the aortic wall were assigned elastic mechanical properties (see Fig. 3.(g)). The element type was turned into S3R in Abaqus®. An isotropic linear elastic material was used for the aortic wall mechanical behavior. The Young's modulus was 2.0 MPa and the Poisson ratio was 0.4 [14]. It was assumed that the aortic behavior could be linearized in the range of relatively small strains induced by the contact with the DB device. All the boundary conditions previously assigned onto the aorta (except the proximal and distal ends which were clamped) were released to establish a static mechanical equilibrium between the aortic wall and the device.

The actual delivery system is introduced through femoral access and a rotation of the sheath during its navigation in the iliac arteries and aorta might happen when the sheath has to slide in tortuous and calcified arteries. As the rotation angle may vary at different longitudinal positions along the sheath, it results into torsion when the device is deployed. After thorough examination of the post-operative CT, we noticed a torsion effect on the second part of the main SG, immediately after the fenestration zone. To mimic the torsion in the simulations, the proximal stent ring was fixed while the distal one was submitted to different degrees (0°, 90°, 135°, 180°) of rotation (see Fig. 1). The other components of the main SG and the BSs were let free.

Approximate placement of Fig 1.

5. Simulation assessment

Positions of stent rings obtained from the post-operative CT scan were chosen as reference positions to validate our deployment simulation. A similar assessment method was used in previous studies [32,33]. Stent rings were manually segmented from CT scans using MevisLab®. Ten points were manually picked on vertebral bones in the CT scans to rigidly register pre- and post-operative CT scans with numerical simulations using the quaternion algorithm method in Matlab®. The mean registration error for the set of 10 points was $2.0 \pm 1.3$ mm.

A first qualitative assessment was achieved by superimposing the deployed stent rings obtained from CT scans and simulations.

A quantitative assessment was also performed for the stent components of the main body. A cylinder was adjusted onto each stent ring using a customized Matlab® routine. For each stent ring, the relative diameter deviation (eD) was estimated between simulations and CT scans. The center position was also assessed by measuring the longitudinal deviation along the arterial centerline (eL) and the transverse deviation in the cross section (eT). Figure 2 shows cylinders matching the post-operative positions of stent rings (a), their simulated positions (b) and the three deviation values (c).

Approximate placement of Fig 2.

6. Sensitivity analyses

- Young modulus

The isotropic linearized elastic model for the aortic wall is a strong simplification so we tested different Young moduli in order to estimate the impact of our assumption. Four values were considered: 1 MPa, 2 MPa, 5 MPa and infinite (rigid arterial wall), in the range of the literature values [14]. A quantitative assessment was then achieved to measure the differences between the four models.

- Friction coefficient

A sensitivity study was conducted to determine whether modifying the friction coefficients between SG components and the aorta could influence the results. Two different friction coefficients were defined: the first one applied on 2cm of the proximal end of the SG (Fprox) and the second one applied on the remaining surface of the SG and on the BGs (Ftot). Six different combinations were tested (0.1-0.2; 0.1-0.4; 0.4-0.2; 0.4-0.4; rough-0.2; rough-0.4), in the range of the literature values [44].

## 5. Results

1. Simulation assessment

Results at different stages of the simulation procedure are shown in Fig. 3.

Deviation values, eD, eL and eT, are shown in Fig. 4 for different torsion angles. Stent rings were numbered from 1 (proximal) to 15 (distal). Stent ring 1 was not included in the quantitative assessment, as it could not be properly segmented from the post-operative CT scan. Stent rings 5, 6 and 7, which were half stents, were not included either. Average deviation values for all stent rings and different torsion angles are reported in Table 1.

Approximate placement of Fig 3.

It can be observed that a 135° torsion produced the best agreement between numerical predictions and post-operative CT. The largest transverse deviations were located in the proximal zone with a maximum eT value equal to 11.4mm. For the longitudinal position, a very good agreement was reached for all stent rings. Regarding diameter deviations, most of the simulated stent rings had the same diameter as the real stent rings (difference less than 6mm) and only one stent ring at the middle showed a larger diameter deviation (error -11.4%).

Approximate placement of Table 1 and Fig 4.

2. Qualitative analysis

The superimposition of simulations and of the post-operative CT scan is shown in Fig. 5. A good agreement between the two configurations was obtained with the 135° torsion model whereas a deviation is clearly visible with the 0° torsion model. Moreover, we observed a satisfactory position of half stents in the fenestration zone for the 135° torsion configuration.

Approximate placement of Fig 5.

3. Sensitivity studies

- Impact of aortic linearized Young's modulus

Position deviations for all stent rings were marginally affected by the Young's modulus of the aortic wall used in our simulation, as shown in Table 2. Mean diameter, longitudinal and transverse deviations ranged within [-10.5% , 1.5%], [-5.8 , -2.3 mm] and [4.4 , 7.2 mm] respectively. We obtained the best global agreement with a 2 MPa modulus.

Approximate placement of Table 2

- Impact of friction coefficients

We tested different levels of friction coefficient in the 0° configuration. Mean deviation values for all stent rings are reported in Table 3. Significant discrepancies in terms of longitudinal deviation were obtained when varying the proximal friction coefficient (Fprox). A proximal defect of apposition was observed with Fprox set to 0.1 or 0.4 and was corrected by a no sliding condition rough contact (see Fig. 6). Overall, the best configuration was obtained with a friction coefficient of 0.4 associated with a rough friction coefficient in the proximal zone.

Approximate placement of Table 3 and Fig 6

6. Discussion

In the present study, we introduced a new methodology to predict computationally DB deployment in patient-specific models of the aortic arch. To the best of our knowledge, this work represents the first report of complex branched SG deployment in the challenging anatomy of the aortic arch using FEA. The presented computational model required the following patient-specific information to be predictive: aortic and SG geometries, aortic wall elastic modulus, friction coefficient between the aortic surface and the SG, position of landing zone and possible torsion angle.

We reported a qualitative and quantitative assessment of our model. We were able to observe a very good agreement between simulations and post-operative CT scans by superimposing the simulated stents onto segmented images of the real stents deduced from post-operative CT scans. We noticed a torsion effect on the post-operative scan, which was well addressed by our simulations with a torsion angle of 135°. Quantitative results were extracted to support the qualitative analysis. Only proximal stent rings showed a rather large transverse deviation, ranging between 8 and 11.4mm. This could be partly explained by the mean registration error of 2.0 ± 1.3mm between pre-operative and post-operative CT scans. Moreover, it should be noticed that images used to analyze the real stent rings were extracted from a non-gated CT scan, which increases the uncertainty at the level of the ascending aorta, due to cardiac pulsation.

We also investigated the effects of torsion on the device. Navigation of the device sheath through tortuous iliac arteries or tortuous aortas can result in a torsion angle between the proximal and distal parts of the sheath. Some cases of torsion have been observed as they induced misalignments of fenestrations during deployment [13]. Due to the length of DB delivery systems used for transfemoral access, the limited ability to precisely control the device end may increase the torsion effect. Torsion induced by guidewire insertion was even simulated by Sanford *et al* [40] in patient-specific models of abdominal aortic aneurysms. As our methodology did not model the navigation, we could not predict the degree of torsion of the SG. An additional boundary condition was assigned to our model to reproduce the torsion and to predict accurately the position of each stent ring after the deployment. Future investigations focused on modeling navigation could be conducted in order to predict the torsion effect in the aortic arch. However, our present study proved successful in reproducing the rotation and showed more specifically that torsion localized in the fenestrated part of the device. This localization phenomenon can be explained by the reduced torsion stiffness of the device in this part with three half stents. It can induce difficulties for catheterizing the internal tunnels of the device through the supra-aortic vessels or even BS kink or carotid arteries coverage, which are major complications. Finally, these numerical simulations could even be used to optimize the design of these complex SGs, for instance by stiffening the torsion response of the SG in the fenestrated region. This may permit to investigate a variety of alternative designs without additional manufacturing costs.

Our model did not take into account the full process of SG deployment during endovascular procedure. It simplified SG crimping, its introduction in a sheath and further navigation through the iliac arteries and aorta. Despite these simplifications, we obtained a very good agreement between the simulation and post-operative CT scans. We reproduced the torsion effect observed in the clinical case and we could successfully predict the effects of this type of complication.

Our model did not consider hemodynamic effects and the fluid-structure interactions [4,6,7,28,36]. Moreover, the action of blood pressure onto the wall was not modeled during simulations as we made the assumption that it already existed in the initial geometry [32]. Future work could refine the action of blood pressure. However, given the level of complexity of our simulations and the very good agreement with post-operative CT scans, quasi-static FEA proved to be a reasonable and reliable first approximation for aortic arch aneurysm, as previously demonstrated in iliac arteries, ascending thoracic aortas and abdominal aortic aneurysms [3,10,11,32–34]. Complementary studies with computational fluid dynamic could help understanding the drag forces acting on the SG and explain further some of the deviations reported in this study [39]. It could also permit to assess the fatigue behavior of the SG under cyclic stresses between diastole and systole with fatigue stress analysis of the stent material [2].

The material model chosen for the vessel wall geometry was an isotropic elastic linearized model and the vessel wall was defined as homogeneous. This is a simplification, which was justified through the sensitivity analysis as we showed that the wall elasticity did not impact significantly the stent positions. The stresses induced by the blood pressure were estimated in previous studies [34], giving indications at which strains and stresses linearized elastic properties of the aorta had to be derived.

Only the extremities of the aortic arch and the supra-aortic trunks were assigned a zero longitudinal displacement during the mechanical equilibrium. Our model did not take into account pressure or tethering by surrounding environment onto the external surface. This effect may probably be pooled with the effect of the aortic stiffness in our model.

Moreover, we disregarded calcifications and the intra-luminal thrombus in the model. They may alter again the mechanical properties of the aortic wall. However, the main contact interactions between the aorta and the SG are located at the proximal and distal landing zone where the intra-luminal thrombus is usually inexistent. Moreover, we calibrated the elastic modulus such that it could take into account the stiffening effect of possible calcifications although they were not visible in the CT scan. Finally, calcifications did not appear in the CT scan of the clinical case considered in our study and the intra-luminal thrombus was very moderate or even inexistent.

We also compared results obtained with a rigid wall and an elastic wall. The rigid model produced larger deviations. Under rigid condition, stent rings were submitted to too strong geometric constraint. Their radial expansion and longitudinal position were too limited. Previous studies dedicated to simulation of SG deployment have considered a variety of mechanical behaviors for the aortic wall, from anisotropic hyperelasticity [17] to orthotropic linearized elasticity [10,33], always showing good agreements with post-operative CT scans. Only recently, Hemmler et al.[19] considered fiber reinforced hyperelastic materials for SG deployment in abdominal aortic aneurysms. In our model, using fiber reinforced hyperelastic materials induced numerical instabilities and high computational cost. Although a good compromise between computational cost and accuracy of our model seemed to have been found, fitting with clinical requirement, future work could consider extending the approach of Hemmler et al.[19] to aortic arch procedures for a more precise prediction of the deformation in the deployed SG. That could also enable stress analyses in the aortic wall after SG deployment, which could be very informative.

Friction coefficients were also investigated through a sensitivity analysis. A significant positive effect was obtained with the rough coefficient (no sliding) in the proximal zone, whereas changing the main coefficient from 0.2 to 0.4 resulted in a modest improvement. This complementary study led us to choose a different friction coefficient depending on the region of the SG: the proximal part had a rough friction contact and the friction coefficient was set to 0.4 elsewhere. This permitted to reach realistic configurations at the proximal zone, avoiding artificial "bird-beak" effects. Away from the proximal region, the friction coefficients assigned to the SG was in agreement with values found in the literature [44]. This sensitivity analysis highlighted the importance of friction in that kind of numerical model.

The obtained values of aortic elastic modulus and friction coefficients may be marginally patient-specific. If similar values can be used for most patients with good accuracy, this would avoid their calibration for every patient, which would simplify the process to render all the computational analyses fully predictive.

The total time to run the simulation was about 48h on 8-cores of the high performance computer of Mines Saint-Etienne (cluster of 11 Tflops with 26 nodes totaling 384 cores and 1 To of RAM). Although this time remains large, it is negligible compared to the duration of the DB custom-made manufacturing process, which lasts about 1 month. Numerical simulation could therefore potentially be included in the planning before the manufacturing process to ensure more reliable device design. It also has the potential to reduce the current manufacturing delay.

The use of endovascular SGs in the aortic arch offers advantages over open surgery. In particular it avoids aortic cross clamping, extracorporeal bypass and consecutive morbidity-mortality. Endovascular procedures permit reducing post-operative risk and shortening hospital stays, which beneficiates mostly to elderly and frail patients [8,15,18,37,41]. But these potential benefits have to be balanced with the lack of long-term feed-back on these relatively new and challenging techniques at the aortic arch level. This study highlights that computational analysis may be used to predict the behavior of complex SG devices in the aortic arch and to detect potential clinical outcomes. Assistance to clinicians through computational simulations could therefore be critical in reducing adverse events and secure this technique in the future.

We demonstrated here the proof of concept of comprehensive and predictive computational simulations of Double Branch Bolton Relay® device deployment and its bridging stents in an aortic arch aneurysm. This work highlights the potential of computational simulations to assist practitioners during pre-operative planning. It also shows the feasibility to simulate, prior to interventions, complete stent-graft deployment despite extremely complex device designs and anatomies. Simulation could not only help in training but also render the planning process faster and more reliable and even assist the practitioners during the procedure. Further studies have to be conducted to apply the methodology on a number of patients to confirm these results.

## 7. Acknowledgement

The authors would like to thank Samuel Arbefeuille, Scott Rush and Christian Fletcher from Terumo Aortic® (formerly Bolton Medical®) for their help and support in this study.

specific numerical simulation of stent-graft deployment: Validation on three clinical cases. *J. Biomech.* 48:1868–1875, 2015.

**List of figures:**

**FIGURE 1. (a)** Stent rings segmented from the post-operative CT scan showing a rotation of the second part of the main stent-graft with respect to the proximal part, manifesting the effects of torsion. **(b)** Simulation approach to model the torsion effect: an additional rotation was applied on the distal stent ring while the two first stent rings were fixed. **(c)** Posterior views of simulation results of the deployment performed with different rotation degrees.

**FIGURE 2. (a)** Cylinders derived from stent ring segmentation for the quantitative assessment approach. The stent ring segmentation was achieved on the post-operative CT scan **(b)** Cylinders derived from stent ring segmentation from the simulation results. **(c)** Schematic definition of deviation measurements between simulations and CT scans.

**FIGURE 3.** Chronological summary of the different steps of the simulation approach. <u>Abbreviations</u>: BSs: bridging stents; SGs: stent-grafts.  <u>Color legend</u>: white: grafts; grey: stents; red: arterial surface.
(a): step 1. SGs modeling and tie constraint highlighted in red
(b): step 2. BSs crimping and placement
(c): step 3. SGs crimping
(d): step 4. BSs bending
(e): step 5. SGs releasing
(f): step 6. SGs deployment
(g): step 7. Mechanical equilibrium
(h): Workflow

**FIGURE 4.** Diameter deviation eD **(a),** longitudinal position deviation eL **(b)** and transverse deviation eT **(c)** depending on the torsion degree (0°, 90°, 135°, 180°). Stent rings are numbered from 1 to 15. Stent rings 1, 5, 6 and 7 are not shown in the comparison. X-axis: stent ring number, Y-axis: deviation value (% or mm)

**FIGURE 5.** Superimposition of the stent rings segmented from the post-operative CT scan and from the simulations. <u>Color legend</u>: grey: post-operative's; blue: torsion 0°; red: torsion 135°. **(a)** Coronal view. **(b)** Transverse view.

**FIGURE 6. (a)** Proximal apposition defect obtained with a friction coefficient of 0.4 for the all stent-graft.  **(b)** Apposition obtained with a rough contact in the proximal region.

**FIGURE 1.** Derycke L. ABME

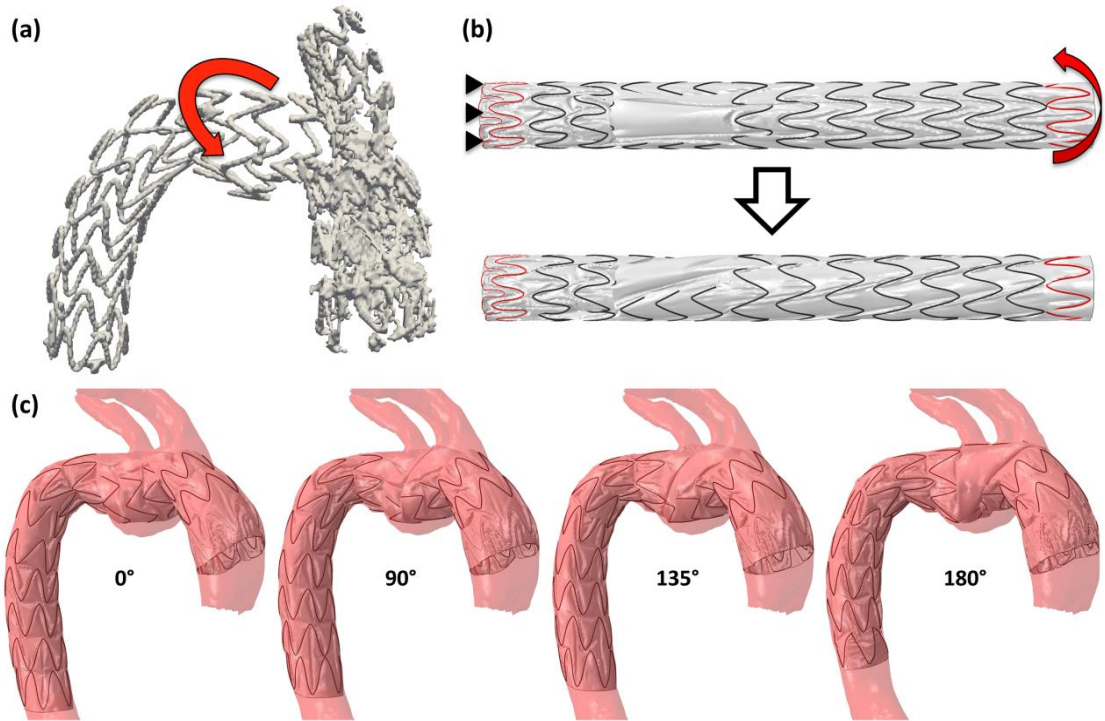

**(a)**

**(b)**

**(c)**

0°          90°          135°          180°

**FIGURE 2.**
Derycke L.
ABME

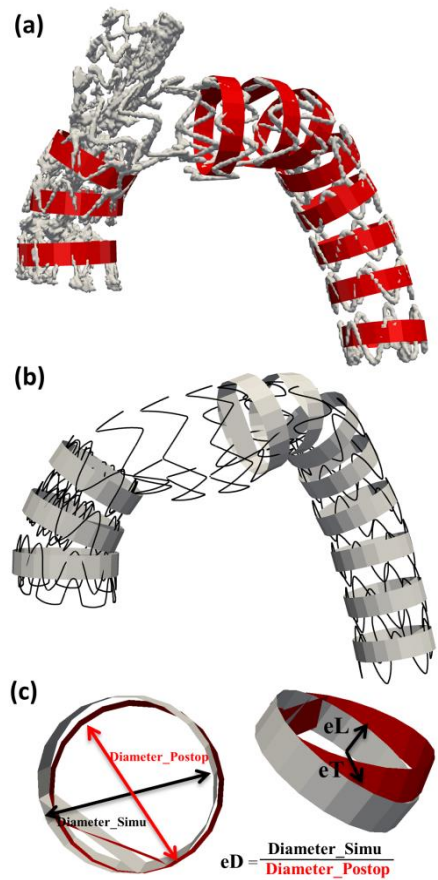

**(a)**

**(b)**

**(c)**

Diameter_Postop

Diameter_Simu

eL

eT

$$eD = \frac{Diameter\_Simu}{Diameter\_Postop}$$

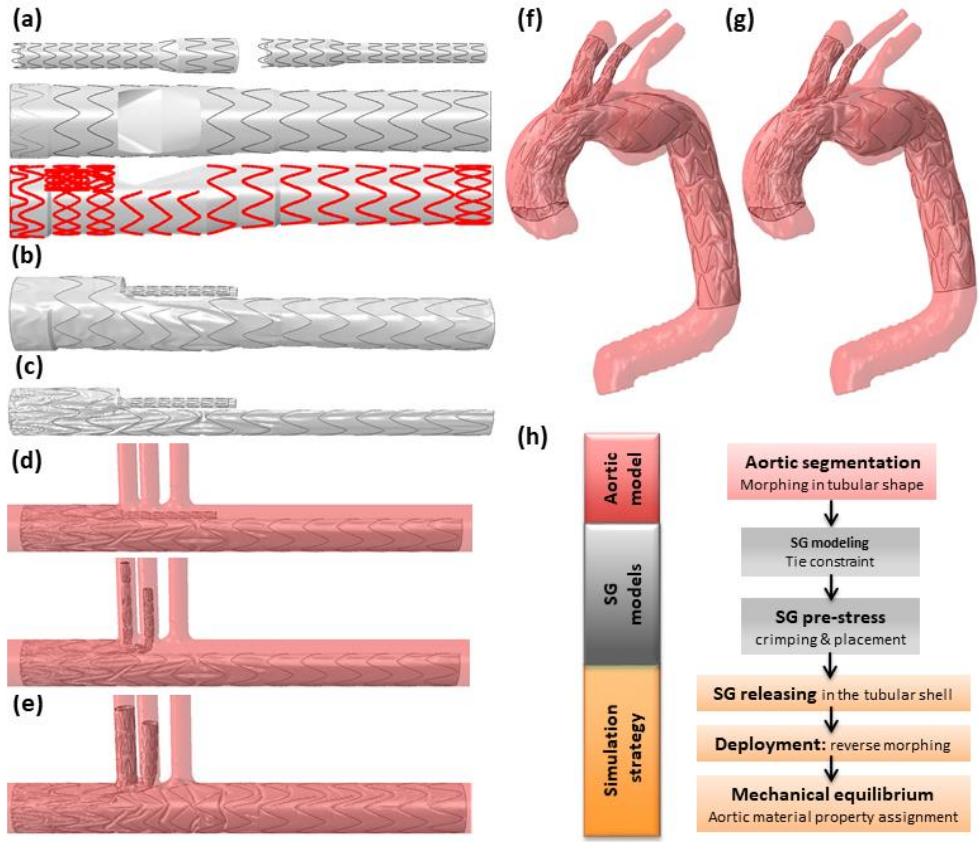

FIGURE 4. Derycke L. ABME

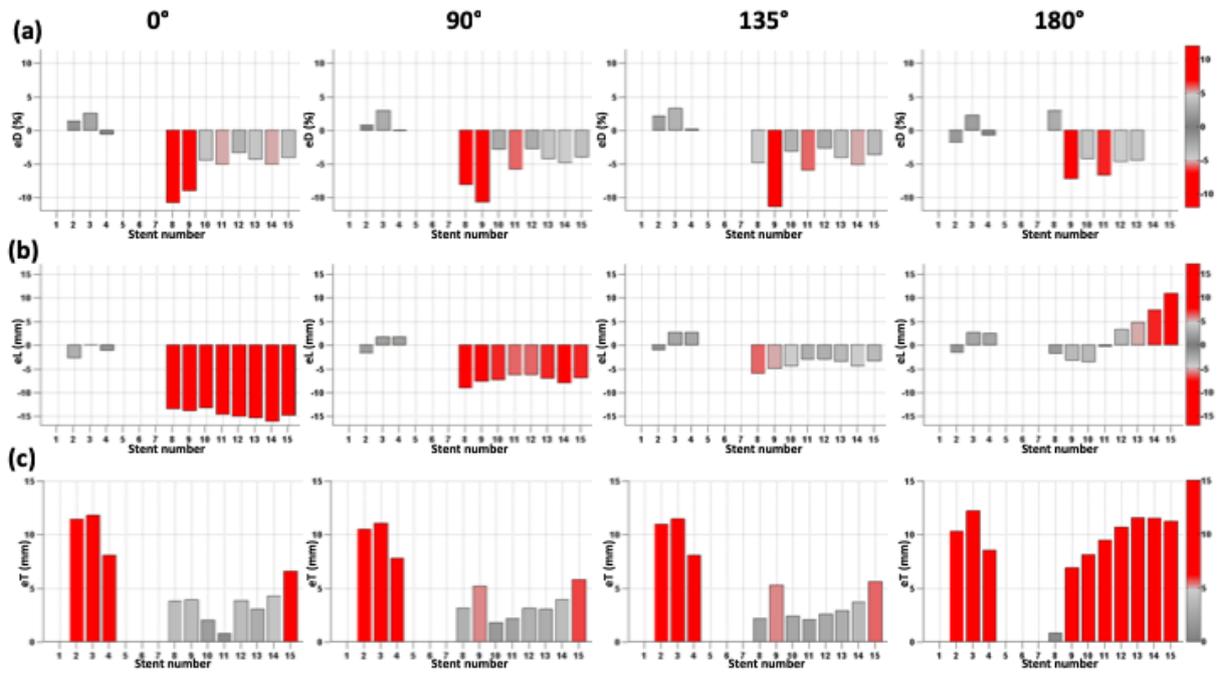

**FIGURE 5.** Derycke L. ABME

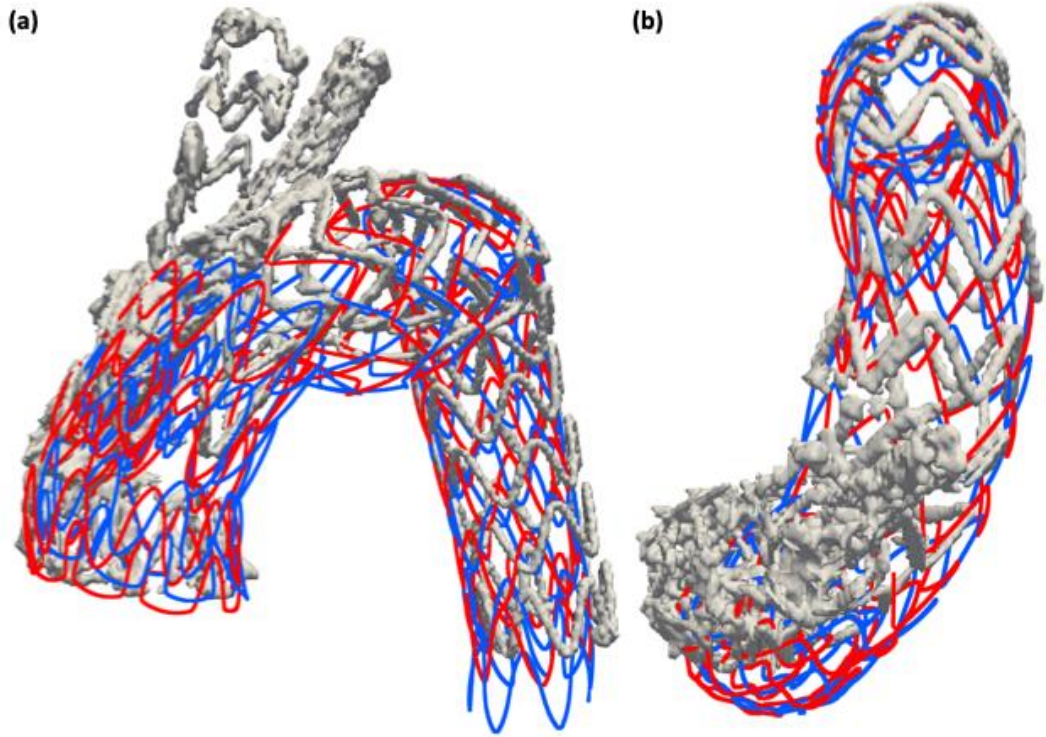

**FIGURE 6.**
Derycke L.
ABME

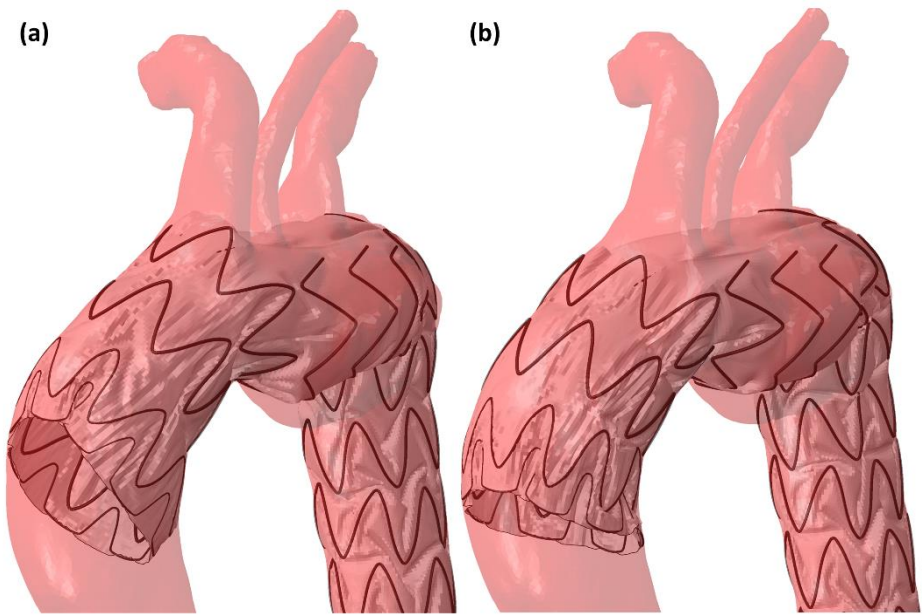

**List of tables:**

**TABLE 1.** Average diameter, longitudinal and transverse deviation values (mean, standard deviation and maximal values) for different values of torsion angle.

**TABLE 2.** Average diameter, longitudinal and transverse deviation values (mean and standard deviation) for different values of Young's modulus.

**TABLE 3.** Average diameter, longitudinal and transverse deviation values (mean and standard deviation) for different values of friction coefficient.

**TABLE 1.**
Derycke L.
ABME

| Torsion (°) | 0 | 90 | 135 | 180 |
|---|---|---|---|---|
| eD (%) | -3.9 ± 4.0 [-10.7 , 2.6] | -3.6 ± 3.9 [-10.7 , 2.9] | -3.2 ± 4.0 [-11.4 , 3.2] | -9.9 ± 16.2 [-41.9 , 2.9] |
| eL (mm) | -11.0 ± 6.3 [-16.1, -0.0] | -5.2 ± 3.9 [-9.1, 1.7] | -2.6 ± 2.9 [-6.0 , 2.6] | 1.9 ± 4.6 [-3.6 , 10.9] |
| eT (mm) | 5.4 ± 3.6 [0.8 , 11.8] | 5.2± 3.2 [1.8 , 11.0] | 5.2 ± 3.5 [2.1 , 11.4] | 9.2 ± 3.2 [0.8 , 12.2] |

**Abbreviations: eD**: relative diameter deviation; **eL**: longitudinal deviation; **eT**: transverse deviation.

**TABLE 2.**
Derycke L.
ABME

| Young modulus | 1 | 2 | 5 | Rigid |
|---|---|---|---|---|
| eD (%) | 1.5 ± 4.8 | -3.2 ± 4.0 | -6.6 ± 3.8 | -10.5 ± 3.8 |
| eL (mm) | -2.3 ± 3.2 | -2.6 ± 2.9 | -4.0 ± 4.6 | -5.8 ± 6.8 |
| eT (mm) | 7.2 ± 4.7 | 5.2 ± 3.5 | 4.5 ± 3.7 | 4.4 ± 2.9 |

**Abbreviations: eD**: relative diameter deviation; **eL**: longitudinal deviation; **eT**: transverse deviation.

**TABLE 3.**
Derycke L.
ABME

| Fprox-Ftot | 0.1–0.2 | 0.1–0.4 | 0.4–0.2 | 0.4–0.4 | Rough–0.2 | Rough–0.4 |
|---|---|---|---|---|---|---|
| eD (%) | -3.2 ± 5.6 | -3.1 ± 5.9 | -3.4 ± 5.1 | -3.2 ± 5.5 | -3.8 ± 4.2 | -3.9 ± 4.0 |
| eL (mm) | -19.1 ± 6.3 | -20.6 ± 6.3 | -16.4 ± 6.2 | -18.5 ± 6.5 | -10.3 ± 6.3 | -11.0 ± 6.3 |
| eT (mm) | 6.6± 4.7 | 7.5 ± 5.1 | 5.8 ± 4.3 | 6.4 ± 4.5 | 5.7 ± 3.8 | 5.4 ± 3.6 |

**Abbreviations: eD**: relative diameter deviation; **eL**: longitudinal deviation; **eT**: transverse deviation.